\documentclass[conference]{IEEEtran}
\usepackage{cite}
\usepackage{bm}
\usepackage{amsmath,amssymb,amsfonts}
\usepackage{algorithmic}
\usepackage{amsthm,amsmath,amssymb}
\usepackage{mathrsfs}
\usepackage{graphicx}
\usepackage{textcomp}
\usepackage{xcolor}
\usepackage{graphicx}
\usepackage{float}
\usepackage{subfigure}
\usepackage{algorithm}

\begin{document}

\title{Latent Semantic Diffusion-based Channel Adaptive De-Noising SemCom for Future 6G Systems}

\author{
\IEEEauthorblockN{Bingxuan Xu\IEEEauthorrefmark{1}, Rui Meng\IEEEauthorrefmark{1}, Yue Chen\IEEEauthorrefmark{1}, Xiaodong Xu\IEEEauthorrefmark{1}\IEEEauthorrefmark{2}, Chen Dong\IEEEauthorrefmark{1}\IEEEauthorrefmark{2}, and Hao Sun\IEEEauthorrefmark{3}}
\IEEEauthorblockA{\IEEEauthorrefmark{1} State Key Laboratory of Networking and Switching Technology, BUPT, Beijing, China}
\IEEEauthorblockA{\IEEEauthorrefmark{2} Department of Broadband Communication, Peng Cheng Laboratory, Shenzhen, China}
\IEEEauthorblockA{\IEEEauthorrefmark{3} Department of Applied Mathematics and Theoretical Physics, University of Cambridge, Cambridge, U. K. }
\IEEEauthorblockA{\{xubingxuan, buptmengrui, mars\_ch, xuxiaodong, dongchen\}@bupt.edu.cn, hs789@cam.ac.uk}}

\maketitle

\begin{abstract}
Compared with the current Shannon's Classical Information Theory (CIT) paradigm, semantic communication (SemCom) has recently attracted more attention, since it aims to transmit the meaning of information rather than bit-by-bit transmission, thus enhancing data transmission efficiency and supporting future human-centric, data-, and resource-intensive intelligent services in 6G systems. Nevertheless, channel noises are common and even serious in 6G-empowered scenarios, limiting the communication performance of SemCom, especially when Signal-to-Noise (SNR) levels during training and deployment stages are different, but training multi-networks to cover the scenario with a broad range of SNRs is computationally inefficient. Hence, we develop a novel De-Noising SemCom (DNSC) framework, where the designed de-noiser module can eliminate noise interference from semantic vectors. Upon the designed DNSC architecture, we further combine adversarial learning, variational autoencoder, and diffusion model to propose the Latent Diffusion DNSC (Latent-Diff DNSC) scheme to realize intelligent online de-noising. During the offline training phase, noises are added to latent semantic vectors in a forward Markov diffusion manner and then are eliminated in a reverse diffusion manner through the posterior distribution approximated by the U-shaped Network (U-Net), where the semantic de-noiser is optimized by maximizing evidence lower bound (ELBO). Such design can model real noisy channel environments with various SNRs and enable to adaptively remove noises from noisy semantic vectors during the online transmission phase. The simulations on open-source image datasets demonstrate the superiority of the proposed Latent-Diff DNSC scheme in PSNR and SSIM over different SNRs than the state-of-the-art schemes, including JPEG, Deep JSCC, and ADJSCC.
\end{abstract}


\section{Introduction}
To support the transition from the Internet of Things (IoT) to the Internet of Everything (IoE), the future sixth-generation (6G) communication systems are expected to enable a wide range of services, including multisensory Extended Reality (XR), connected robotics and autonomous systems, wireless brain-computer interactions, and more. This will be achieved through emerging techniques such as higher millimeter wave (mmWave) frequencies, large intelligent surfaces, edge Artificial Intelligent (AI), and integrated terrestrial, airborne, and satellite networks \cite{saad2019vision}. However, the current Shannon’s Classical Information Theory (CIT) paradigm faces several challenges in supporting human-centric, data-, and resource-intensive intelligent services. These challenges include wireless transmission of a large amount of data, rapid system response as well as reliable and efficient information interaction, and consumption of more network resources for real-time updating of information and analysis of user data \cite{yang2022semantic}.

In response, semantic communication (SemCom) is a promising revolutionary paradigm in 6G to breakout the “Shannon’s trap” through filtering redundant information and extracting the meaning of effective information. Owing to the development of distributed computation and wide connectivity of ubiquitous intelligent devices, SemCom is expected to be deployed on a large scale in 6G. Particularly, SemCom has the following superiorities, including relieving the pressure of data transmission, improving network management efficiency, and enhancing resource allocation effectiveness \cite{yang2022semantic2}\cite{dong2022semantic}.

Thanks to the growing up of Deep Learning (DL) techniques in comprehending language and audio as well as pictures, DL-based communication architectures have been recently developed. O’Shea \textit{et al.} \cite{o2016learning} design an Auto-Encoders (AEs)-based end-to-end wireless communication system, where the channel AEs include the encoder, channel regularizer, and decoder, and convolutional layers and domain specific regularizing effects are introduced to reduce the number of parameters and avoid over-fitting issues, respectively. O’Shea \textit{et al.} \cite{o2016learning} further verify that the proposed system not only has comparable performance potential with modern systems, but also almost reaches Shannon capacity, while maintaining universality and low complexity.

More recently, the joint source-channel coding (JSCC) schemes for structured sources in SemCom have attracted more attention, since such design is more feasible for actual communication systems with constrained block lengths. Xie \textit{et al.} \cite{xie2021deep} present a SemCom framework based on Transformer and transfer learning to extract the semantic information of texts during noisy environments. Xie \textit{et al.} \cite{xie2021deep} also consider cross-entropy and mutual information in the loss function to maximize the system capacity and propose a novel metric to measure the semantic error of sentences. Weng \textit{et al.} \cite{weng2021semantic} exploit squeeze-and-excitation (SE) networks to learn essential speech semantic information and employ the attention mechanism to enhance the accuracy of signal recovery. Bourtsoulatze \textit{et al.} \cite{bourtsoulatze2019deep} first develop a SemCom architecture based on Convolution Neural Networks (CNNs) for radio transmission of high-resolution pictures over additive white Gaussian noise (AWGN) and fading channels. Dong \textit{et al.} first design semantic slice-models to adaptively realize semantic transmission tasks in different circumstances \cite{dong2022semantic}. Du \textit{et al.}  \cite{du2023ai} propose an AI-generated incentive mechanism based on contract theory to facilitate semantic information sharing among users. The diffusion model is utilized to generate the optimal contract design, which effectively promote the sharing of semantic information between users.

Although many researchers pay attention to DL-enabled SemCom and utilize its potential in enhancing communication efficiency, SemCom is still facing some challenges. For example, channel noises severely affect the recovery and reception of semantic information. Hence, suppressing channel noises in the restoration of semantic information should be appropriately addressed to enhance transmission performance.

To resist channel noises, the schemes proposed in \cite{xie2021deep} and \cite{bourtsoulatze2019deep} introduce channels when training models to recover signals at the specific SNR. However, the above-mentioned schemes do not consider that there may exist differences between the SNR levels during the training and deployment stages. On the other hand, it is computationally inefficient for multi-networks to cover the scene with a wide range of SNRs. To realize highly robust SemCom during different SNRs, Hu \textit{et al.} \cite{hu2022robust} develop a vector quantization-based semantic communication system, which utilizes adversarial networks to perform semantic de-noising for image classification tasks. For image generation, Xu \textit{et al.} \cite{xu2021wireless} propose an Attention DL-based JSCC (ADJSCC) scheme by the attention mechanism to dynamically adjust SNRs during the training phase. Such design can capture inherent channel characteristics at different SNRs and remove channel noises during actual deployment. Nevertheless, adaptive de-noising in different SNR environments remains a major challenge.

To support highly reliable image transmission for future 6G systems, we combine Variational Autoencoder (VAE), adversarial learning, and diffusion model to realize de-noising semantic communication under noisy channel environments. The main contributions are described as follows:
\begin{itemize}
\item{We develop the De-Noising SemCom (DNSC) framework consisting of the encoder and decoder to achieve highly robust transmission for future 6G systems, where the proposed semantic de-noiser module in the decoder can relieve the effects of channel noises on semantic transmission.}
\item{To realize adaptive online semantic de-noising under various SNR environments, we further propose the Latent Diffusion DNSC (Latent-Diff DNSC) scheme upon the designed DNSC system. The constructed objective function of the encoder and decoder joint training considers the reconstruction loss and adversarial loss as well as regularization loss for optimizing VAEs, optimizing the discriminator that differentiate original and generated images, and avoiding arbitrarily scaled latent semantic spaces, respectively.}
\item{The semantic de-noiser in the proposed Latent-Diff DNSC scheme is obtained through forward and reverse diffusion processes. The gradually added noises in the forward phase to model real channel noises are iteratively eliminated in the reverse phase by maximizing the logarithmic likelihood of the distribution predicted by the DL model. Under such design, the channel noises under different SNR conditions can be adaptively removed in the online inference stage without knowing channel information in the offline training stage.}
\item{The simulation results on open-source LAION2B-EN dataset \cite{schuhmann2022laion} demonstrate that, under the same compression ratio, the proposed Latent-Diff DNSC model achieves higher Peak Signal-to-Noise Ratio (PSNR) by $20\%\sim67\%$ and higher Structural SIMilarit (SSIM) by $4\%\sim68\%$ compared to the ADJSCC and JSCC model.}
\end{itemize}

\section{System Model}

\begin{figure}[t]
\centering
\includegraphics[width=0.48\textwidth]{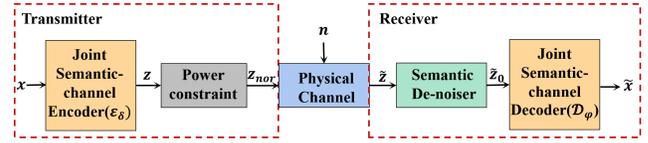}
\caption{Illustration of the proposed DNSC architecture.}
\label{mo}
\vspace{-1em}
\end{figure}

As shown in Fig. \ref{mo}, a typical SemCom architecture is considered, including the transmitter, physical channels, and the receiver, which are described as follows in detail.

\subsection{Transmitter}
The input data $\mathbf{x}$ undergoes joint semantic-channel encoder, leading to the formation of a semantic latent vector $\mathbf{z}$.
\begin{equation}
\mathbf{z} =\mathcal{E}_{\delta}(\mathbf{x})\label{en}
\end{equation}

As shown in (\ref{power}), the transmitted signal is subject to a power constraint of magnitude $P$ before being transmitted through the physical channel. 
\begin{equation}
\mathbf{z}_{nor}=\sqrt{k P} \frac{\mathbf{z}}{\sqrt{\mathbf{z}\mathbf{z}^*}}\label{power}
\end{equation}

\subsection{Physical Channel}
The physical channel is usually modeled as the additive white Gaussian noise (AWGN), denoted by $\boldsymbol{n}\sim \mathcal{CN}(0,\sigma_w^2)$, where $\sigma_w^2$ represents the power of the noise. Hence, the received signal $\mathbf{\tilde{z}}$ at the receiver is denoted as
\begin{equation}
\mathbf{\tilde{z}} = \boldsymbol{h}*\mathbf{z}_{nor}+\boldsymbol{n}\label{channel}
\end{equation}
where $\boldsymbol{h}$ represents the coefficients of the physical channel between the transmitter and receiver.

\subsection{Receiver}

\emph{Definition 1 (Semantic De-Noiser):}
Unlike the existing SemCom frameworks, in the proposed DNSC architecture, the semantic de-noiser module at the receiver is defined to remove noise components from the noisy semantic vector. Such design can be achieved by neural networks parameterized by $\theta$. 

By the prediction of the trained semantic de-noiser, latent semantic variables $\mathbf{\tilde{z}}$ are progressively denoised and eventually transformed into $\mathbf{\tilde{z}}_0$. Then, the de-noised semantic vector $\mathbf{\tilde{z}}_0$ is sent to the joint semantic-channel decoder to generate $\mathbf{\tilde{x}}$ as
\begin{equation}
    \mathbf{\tilde{x}} = \mathcal{D}_{\varphi}(\mathbf{\tilde{z}}_0)\label{de}
\end{equation}

 Overall, the proposed semantic de-noiser allows it to accurately model noise characteristics of physical channels, leading to effective de-noising of the semantic information. In this way, the proposed DNSC architecture can intelligently remove noises from the latent semantic vectors to improve the robustness and generalization ability of SemCom models. The specific implementation of the semantic de-noiser will be provided in Section III.

\section{The Proposed Latent-Diff DNSC system}
\begin{figure}
\centering
\subfigure[ ]{
\includegraphics[width=0.43\textwidth,height=0.3\textwidth]{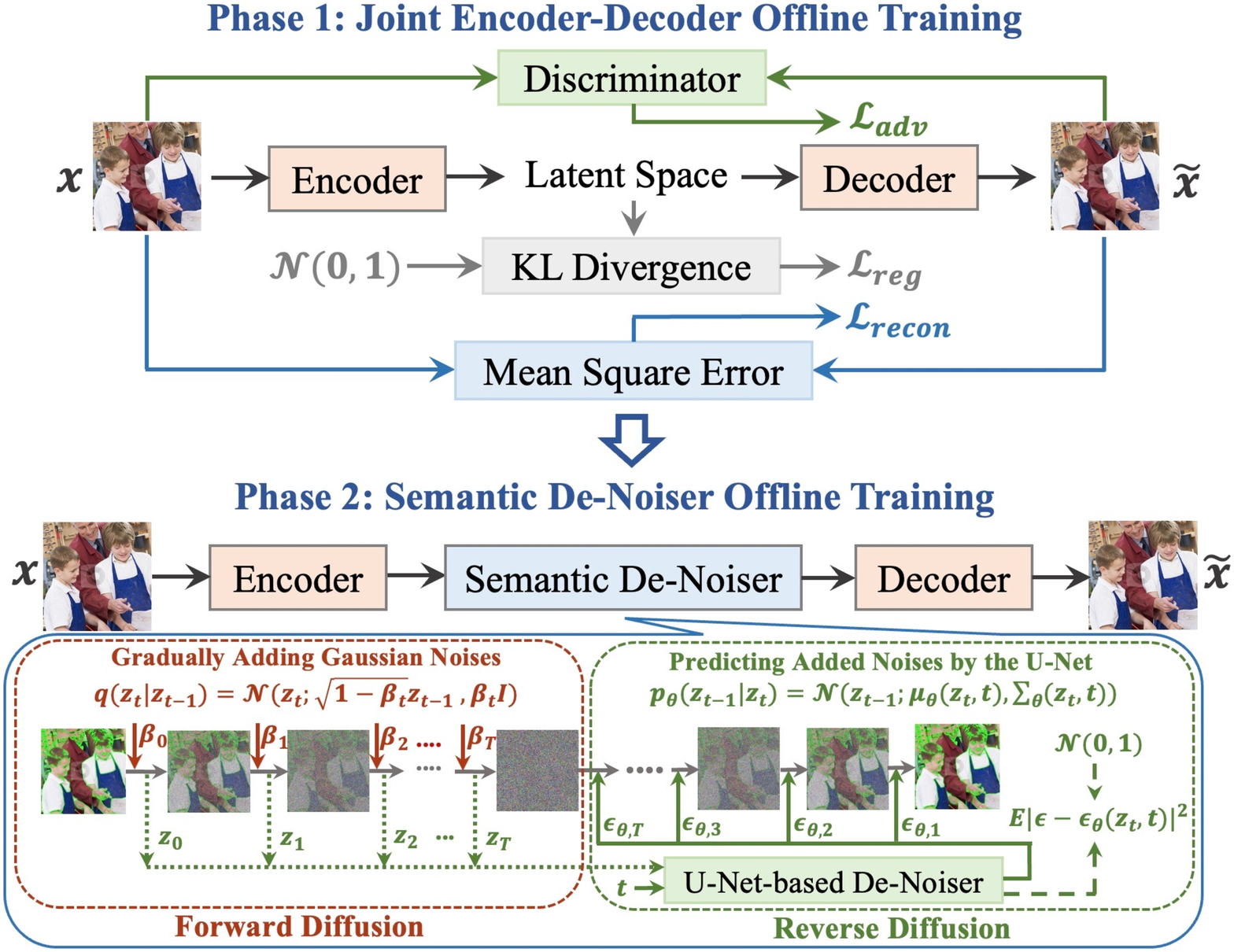} 
}
\subfigure[ ]{
\includegraphics[width=0.43\textwidth,height=0.09\textwidth]{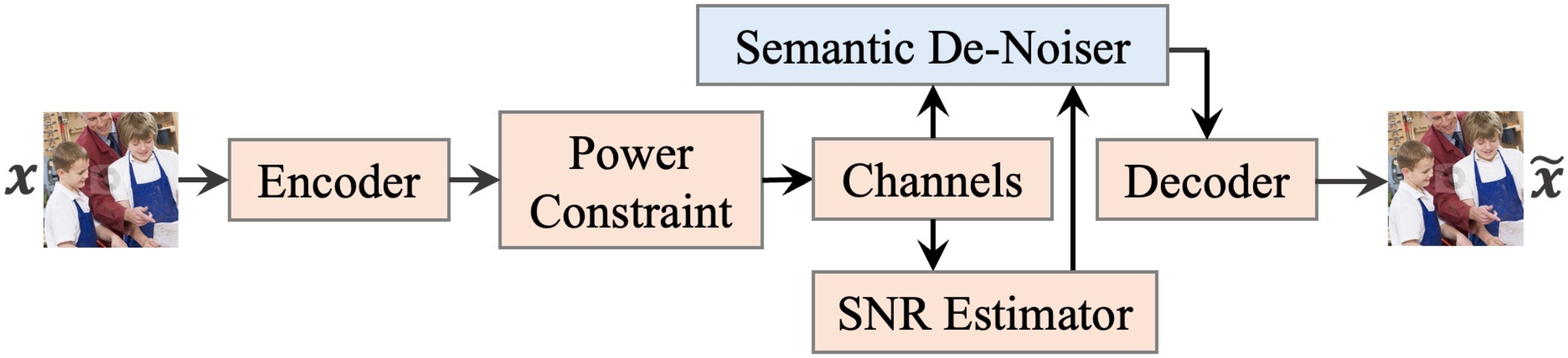} 
}
\DeclareGraphicsExtensions.
\caption{Illustration of the proposed Latent-Diff DNSC model. (a): offline training; (b): online inference.}
\label{model}
\vspace{-1em}
\end{figure}

As illustrated in Fig. \ref{model}, upon the designed DNSC system, we present the Latent-Diff DNSC scheme, including the offline training and online inference phases.

\addtolength{\topmargin}{0.011in}
\subsection{Offline Training phase}\label{AA}
As illustrated in Algorithm 1, the offline training phase includes two training stages.
\begin{algorithm}[!h]
\caption{The Offline Training Algorithm for Latent-Diff DNSC}
\hspace*{0.02in} {\textbf{The first training stage}}
\begin{algorithmic}[1]
\STATE \textbf{Input:} Training dataset $\mathcal{X}$, batch size $N$, learning rate $\gamma$;
\REPEAT
\STATE Sample a data batch $\mathbf{x} = [\boldsymbol{x}_1,...,\boldsymbol{x}_N]$ from $\mathcal{X}$;
\FOR{each data sample $\boldsymbol{x}_i \in \mathbb{R}^{H\times W \times 3}$ in $\boldsymbol{X}$}
\STATE The encoder $\mathcal{E}_{\delta}$ encodes $\boldsymbol{x}_i$ into a latent representation $\boldsymbol{z}_i\in \mathbb{R}^{h\times w \times c}$,$\boldsymbol{z_i}=\mathcal{E}_{\delta}(\boldsymbol{x}_i)$;
\STATE The decoder $\mathcal{D}_{\varphi}$ reconstructs image $\boldsymbol{\tilde{x}}_i$ from latent $\boldsymbol{z}_i$, $\boldsymbol{\tilde{x}}_i=\mathcal{D}_{\varphi}(\boldsymbol{z}_i)=\mathcal{D}_{\varphi}(\mathcal{E}_{\delta}(\boldsymbol{x}_i))$;
\ENDFOR
\STATE Compute the mean $\mu_{z}$ and variance $\sigma_{z}$ of the latent $\mathbf{z}$;\\
\STATE Compute reconstruction loss via (\ref{lrecon});\\
\STATE Compute adversarial loss via (\ref{adv});\\
\STATE Compute regularization loss via (\ref{reg});\\
\STATE  Compute average loss via (\ref{l1});\\
\STATE Update the parameter $\delta$ of the encoder $\mathcal{E}$, $\varphi$ of the decoder $\mathcal{D}$ and the parameter $\omega$ of the discriminator $D$ by gradient descent;
\UNTIL {Converged}
\end{algorithmic}

\hspace*{0.02in} {\textbf{The second training stage}}
\begin{algorithmic}[1]
\STATE \textbf{Input:} hyperparameter of Gaussian distribution variance $\{\beta_1,\beta_2,...\beta_T\}$;
\REPEAT
\STATE Sample a batch of latent representation $\mathbf{z}$ from the first stage, $\mathbf{z}_{0}\sim q(\mathbf{z}_{0})$; 
\STATE $t\sim$ Uniform $({1,...,T})$;
\STATE $\boldsymbol{\epsilon} \sim \mathcal{N}(0,\mathbf{I})$;
\STATE Compute $\alpha_t = 1-\beta_t$ and  $\bar{\alpha}_t = \alpha_1\times\alpha_2\times ...\alpha_t $;
\STATE Take gradient descent step on (\ref{l2});
\UNTIL {Converged}
\end{algorithmic}
\end{algorithm}

\subsubsection{The First Training Stage}
During the initial phase, the encoder $\mathcal{E}_{\delta}$ and decoder $\mathcal{D}_{\varphi}$ are jointly trained to obtain the distribution of the semantic latent space. As illustrated in Table \ref{tab:encoder-decoder}, the encoder-decoder structure in \cite{rombach2022high} is adopted.

\emph{Proposition 1:} After the joint training, to learn the probability distribution of the latent semantic space, the model is trained by the constructed loss function involved the reconstruction loss $\mathcal{L}_{recon}$, self-similarity loss $\mathcal{L}_{adv}$, and regularized Kullback-Leibler (KL) divergence loss $\mathcal{L}_{reg}$ as
\begin{equation}
\mathcal{L}_1=\mathcal{L}_{\text {recon}}+\lambda_{\text {adv}} \mathcal{L}_{\text{adv}}+\lambda_{\text {reg }} \mathcal{L}_{\text {reg}}\label{l1}
\end{equation}
where $\lambda_{adv}$ and $\lambda_{reg}$ denote the weights of the loss function, $\mathcal{L}_{recon}$ quantifies the disparity between the original and reconstructed images, $\mathcal{L}_{adv}$ obtained through adversarial training preserves local consistency, and $\mathcal{L}_{reg}$ computed in the latent space encourages the retention of maximal information while encoding images, which can be denoted as

\begin{equation}
\mathcal{L}_{\text {recon}}=|\mathbf{\tilde{x}}-\mathbf{x}\|_{2}^{2}\label{lrecon}
\end{equation}

\begin{equation}
\mathcal{L}_{a d v}=[\log {D_{\omega}}(\mathbf{x})]+[\log(1-{D}_{\omega}(\mathbf{\tilde{x}}))])\label{adv}
\end{equation}
where $D_{\omega}$ is the discriminator of the parameter $\omega$.
\begin{equation}
\mathcal{L}_{\text {reg }}=KL(\mathcal{N}(\mu_{z}, \sigma_{z}) \| \mathcal{N}(0,1))\label{reg}
\end{equation}

\begin{table}[t]
    \centering
    \caption{The structure of Encoder-Decoder}
    \label{tab:encoder-decoder}
    \begin{tabular}{@{}l@{\hspace{0.8pt}}l@{}}
        \hline
        \fontsize{8}{9}\selectfont{\textbf{Encoder}} & \fontsize{8}{9}\selectfont{\textbf{Decoder}}\\
        \hline
        \fontsize{8}{9}\selectfont{$\mathbf{x}\in \mathbb{R}^{H \times W \times C}$} & \fontsize{8}{9}\selectfont{$\mathbf{z}\in \mathbb{R}^{h \times w \times c}$ }\\
        \hline
        \fontsize{8}{9}\selectfont{Conv2D$:  \mathbb{R}^{H \times W \times C^{\prime}}$} & \fontsize{8}{9}\selectfont{Conv2D$: \mathbb{R}^{H \times W \times C^{\prime \prime}}$}\\
        \hline
        \fontsize{7}{8}\selectfont{$m\times{\text {ResBlok,Downsample}}: \mathbb{R}^{h \times w\times C^{\prime \prime}}$} & \fontsize{8}{9}\selectfont{ResBlok$: \mathbb{R}^{h\times w\times C^{\prime \prime}}$}\\
        \hline
        \fontsize{8}{9}\selectfont{ResBlok $: \mathbb{R}^{h\times w\times C^{\prime \prime}}$} &\fontsize{8}{9}\selectfont{ Non-Local$: \mathbb{R}^{h\times w\times C^{\prime\prime}}$ }\\
        \hline
        \fontsize{8}{9}\selectfont{Non-Local$:  \mathbb{R}^{h\times w\times C^{\prime \prime}}$} & \fontsize{8}{9}\selectfont{ResBlok$: \mathbb{R}^{h\times w\times C^{\prime \prime}}$} \\
        \hline
        \fontsize{8}{9}\selectfont{ResBlok$:  \mathbb{R}^{h\times w\times C^{\prime \prime}}$}& \fontsize{7}{8}\selectfont{$m \times{\text{ResBlok,Upsample}}: \mathbb{R}^{H\times W\times C^{\prime}}$}\\
        \hline
        \fontsize{8}{9}\selectfont{GroupNorm,Swish,Conv2D$: \mathbb{R}^{h\times w\times c}$}&\fontsize{7}{8}\selectfont{ GroupNorm,Swish,Conv2D$: \mathbb{R}^{H\times W\times C}$}\\
        \hline
    \end{tabular}
    \footnotesize 
\end{table}

\subsubsection{The Second Training Stage}

A forward diffusion chain progressively adds noise to the semantic latent vectors $\mathbf{z}_0 = \mathcal{E}_{\alpha}(\mathbf{x})$ via a Markov chain, using an ascending variance schedule ${\beta_1,\beta_2,..., \beta_T}$ as
\begin{equation}
q\left(\mathbf{z}_{1: T} \mid \mathbf{z}_{0}\right)=\prod_{t=1}^{T} q\left(\mathbf{z}_{t} \mid \mathbf{z}_{t-1}\right)
\end{equation}
where 
\begin{equation}
q\left(\mathbf{z}_{t} \mid \mathbf{z}_{t-1}\right)=\mathcal{N}\left(\mathbf{z}_{t}; \sqrt{1-\beta_{t}} \mathbf{z}_{t-1}, \beta_{t} \mathbf{I}\right)
\end{equation}
After reparameterization, $\mathbf{z}_{t}$ can be get by
\begin{equation}
\mathbf{z}_{t}=\sqrt{\bar{\alpha}_{t}} \mathbf{z}_{0}+\sqrt{1-\bar{\alpha}_{t}} \mathbf{\epsilon}\label{reor}
\end{equation}
where $\bar{\alpha}_{t}=\alpha_1\times \alpha_2 \times ...\alpha_t$, $\alpha_t = 1-\beta_t$.

The training data is transformed into semantic vectors $\mathbf{z}_{0}$ with a distribution of $q(\mathbf{z}_{0})$ through the trained encoder. Gaussian noises are also added to model real-world noise conditions. $q(\mathbf{z}_{0})$  is iteratively added with noise for $T$ steps and gradually becomes a Gaussian noise distribution $\boldsymbol{\epsilon}\sim q(\mathbf{z}_{T})$. In the reverse process, our goal is to train a U-shaped Network (U-Net) with parameters $\theta$ to learn the inverse distribution $p_{\theta}$, which enables us to gradually recover the original latent distribution of $\mathbf{z}_{0}$ from the diffused latent $\mathbf{z}_{T}$ through the reverse diffusion process. The reverse process is given by
\begin{equation}
p_{\theta}\left(\mathbf{z}_{0}\right)=\int_{z} p_{\theta}\left(\mathbf{z}_{T}\right) \prod_{t=1}^{T} p_{\theta}\left(\mathbf{z}_{t-1} \mid \mathbf{z}_{t}\right)\label{p}
\end{equation}
where $p_{\theta}\left(\mathbf{z}_{t-1} \mid \mathbf{z}_{t}\right)=\mathcal{N}\left(\mathbf{z}_{t-1} ; \mu_{\theta}\left(\mathbf{z}_{t}, t\right), \Sigma_{\theta}\left(\mathbf{z}_{t}, t\right)\right)$. Although it is difficult to obtain the probability distribution $q(\mathbf{z}_{t-1}|\mathbf{z}_t)$ of the reverse process, $q(\mathbf{z}_{t-1}|\mathbf{z}_t, \mathbf{z}_0)$ can be obtained by (\ref{xxx}) if $\mathbf{z}_0$ is known.
\begin{equation}
    q\left(\mathbf{z}_{t-1} \mid \mathbf{z}_{t}, \mathbf{z}_{0}\right)=\mathcal{N}\left(\mathbf{z}_{t-1} ; \tilde{\mu}\left(\mathbf{z}_{t}, \mathbf{z}_{0}\right), \bar{\beta}_{t} \mathbf{I}\right)\label{xxx}
\end{equation}
where ${\bar{\beta}}_t = {\beta}_1\times{\beta}_2\times...{\beta}_t$. The parameters can be optimized by maximizing the evidence lower bound (ELBO) as
\begin{equation}
\begin{aligned}
\log p_{\theta}\left(\mathbf{z}_{0}\right) &\geq -\mathbb{K} \mathbb{L}\left(q\left(\mathbf{z}_{T} \mid \mathbf{z}_{0}\right) \mid p\left(\mathbf{z}_{T}\right)\right)-\\&\sum_{t=1}^{T} \mathbb{E}_{q\left(\mathbf{z}_{t} \mid \mathbf{z}_{0}\right)} \mathbb{K} \mathbb{L}\left(q\left(\mathbf{z}_{t-1} \mid \mathbf{z}_{t}, \mathbf{z}_{0}\right) \mid p\left(\mathbf{z}_{t-1} \mid \mathbf{z}_{t}\right)\right)
\end{aligned}
\end{equation}

\emph{Proposition 2:} Essentially, the proposed semantic de-noiser is composed of a set of neural networks $\boldsymbol{\epsilon}_{\theta}(\mathbf{z}_{t}, t)$ with similar parameters and weights, and is trained to estimate the added noise variables for input $\mathbf{z}_t$. 
Besides, considering that $\mathbf{z}_0$ is unknown, $\mathbf{z}_0$ is replaced by transforming (\ref{reor}), we further obtain the loss function as (\ref{l2}). 

\begin{equation}
\mathcal{L}_2 =\mathbb{E}_{\mathcal{E}_{\alpha}(\mathbf{x}), \boldsymbol{\epsilon} \sim \mathcal{N}(0,1), t}\left[\left\|\boldsymbol{\epsilon}-\boldsymbol{\epsilon}_{\theta}\left(\mathbf{z}_{t}, t\right)\right\|_{2}^{2}\right]\label{l2}
\end{equation}

Hence, the semantic vector $\mathbf{z}_t$ is the input of the semantic de-noiser at time $t$, and the corresponding noise variables are estimated and outputted. After the training process gradually converges by minimizing the loss function in (\ref{l2}), the trained semantic de-noiser can accurately eliminate noises.
\addtolength{\topmargin}{0.011in}
\subsection{Online inference phase}
The online inference process is illustrated in Algorithm 2. Adding noises in the training phase can result in a higher overall level of noises, which can negatively impact the transmission accuracy. To mitigate this problem, power normalization is employed to ensure that the variance of noises is proportional to the power of signals, leading to a final noise power level of $1$ after $T$ steps of diffusion. Therefore, to eliminate semantic noise effectively during the inference phase, it is necessary to obtain the normalization factor $\phi$ at the receiver through SNR estimation, thus normalizing the power of the received signal as
\begin{equation}
\mathbf{\tilde{z}}_T=\phi\mathbf{\tilde{z}}=\frac{\mathbf{\tilde{z}}}{\sqrt{(1/10^{SNR/10})}\sqrt{\tilde{\mathbf{z}}\tilde{\mathbf{z}}^*}}\label{nor}
\end{equation}
where $\boldsymbol{\tilde{z}}$ is the semantic vector transmitted through the channel in (\ref{channel}). Subsequently, the signal $\boldsymbol{\tilde{z}}_T$ is fed into de-noiser for noise elimation. After $T$ iterations, the final output is the recovered semantic vector $\boldsymbol{z}_0$, which is then input into the decoder to obtain $\boldsymbol{\tilde{x}}$.

\begin{algorithm}[t]
\caption{The Online Inference Algorithm for Latent-Diff DNSC}
\begin{algorithmic}[1]
\STATE \textbf{Input:} Image data $\mathbf{x}$ to be transmitted, hyperparameter of Gaussian distribution variance $\{\beta_1,\beta_2,...\beta_T\}$;
\STATE The encoder $\mathcal{E}_{\delta}$ encodes $\mathbf{x}$ into a latent representation $\mathbf{z}$ via (\ref{en});
\STATE Power constraints via (\ref{power});
\STATE $\mathbf{z}$ undergoes channel transmission and becomes $\mathbf{\tilde{z}}$ at the receiving end via (\ref{channel});
\STATE The SNR estimation module receives $\mathbf{\tilde{z}}$ and normalizes the power via (\ref{nor});
\FOR{de-noising steps = T,...,1}
\STATE $\mathbf{y}\sim\mathcal{N}(0,\mathbf{I})$ if $t\ge1$,else $\mathbf{y} = 0$;
\STATE $\mathbf{\tilde{z}}_{t-1}=\frac{1}{\sqrt{\alpha_{t}}}\left(\mathbf{\tilde{z}}_{t}-\frac{1-\alpha_t}{\sqrt{1-\bar{\alpha}_{t}}} \boldsymbol{\epsilon}_{\theta}\left(\mathbf{\tilde{z}}_{t}, t\right)\right)+\sigma_{t} \mathbf{y}$;
\ENDFOR

\STATE The decoder $\mathcal{D}_{\varphi}$ reconstructs image $\mathbf{\tilde{x}}$ from the received signal $\mathbf{\tilde{z}}$ via (\ref{de});

\end{algorithmic}
\end{algorithm}

\section{Simulation Results and Analysis}
In this section, we evaluate the performance of the proposed Latent-Diff DNSC scheme on AWGN channel with perfect SNR estimation. The proposed model is pre-trained by \cite{rombach2022high}. The diffusion steps $T$ for the pre-trained model is $1000$, the learning rate is $9.6\times 10^{-5}$, the training batchsize is $8$, and the noise schedule is linear with the increase in $\{\beta_1,\beta_2,...\beta_T\}$.
\subsection{Open-Source Image Dataset}
We select 660 images from the LAION2B-EN dataset \cite{schuhmann2022laion} to compare the transmission performance of the proposed scheme with baseline schemes. The LAION2B-EN dataset is an image-text dataset, and all images are resized to have the shape of $3 \times 512 \times 512$, where the dimensions correspond to RGB channels, width, and height, respectively.
\subsection{Baseline Schemes}
We compare the performance of our proposed scheme with DeepJSCC\cite{bourtsoulatze2019deep}, ADJSCC\cite{xu2021wireless}, and the traditional JPEG image compression algorithm. DeepJSCC is trained with two SNRs, including 10 dB and 20 dB, ADJSCC is trained with SNRs ranging from 0-28 dB. The Latent-Diff DNSC scheme without the de-noiser module is also provided as a baseline scheme to highlight the significance of the de-noiser.
\subsection{Performance Comparison}
\begin{figure}[t]
\centering
\includegraphics[width=0.45\textwidth,height=0.3\textwidth]{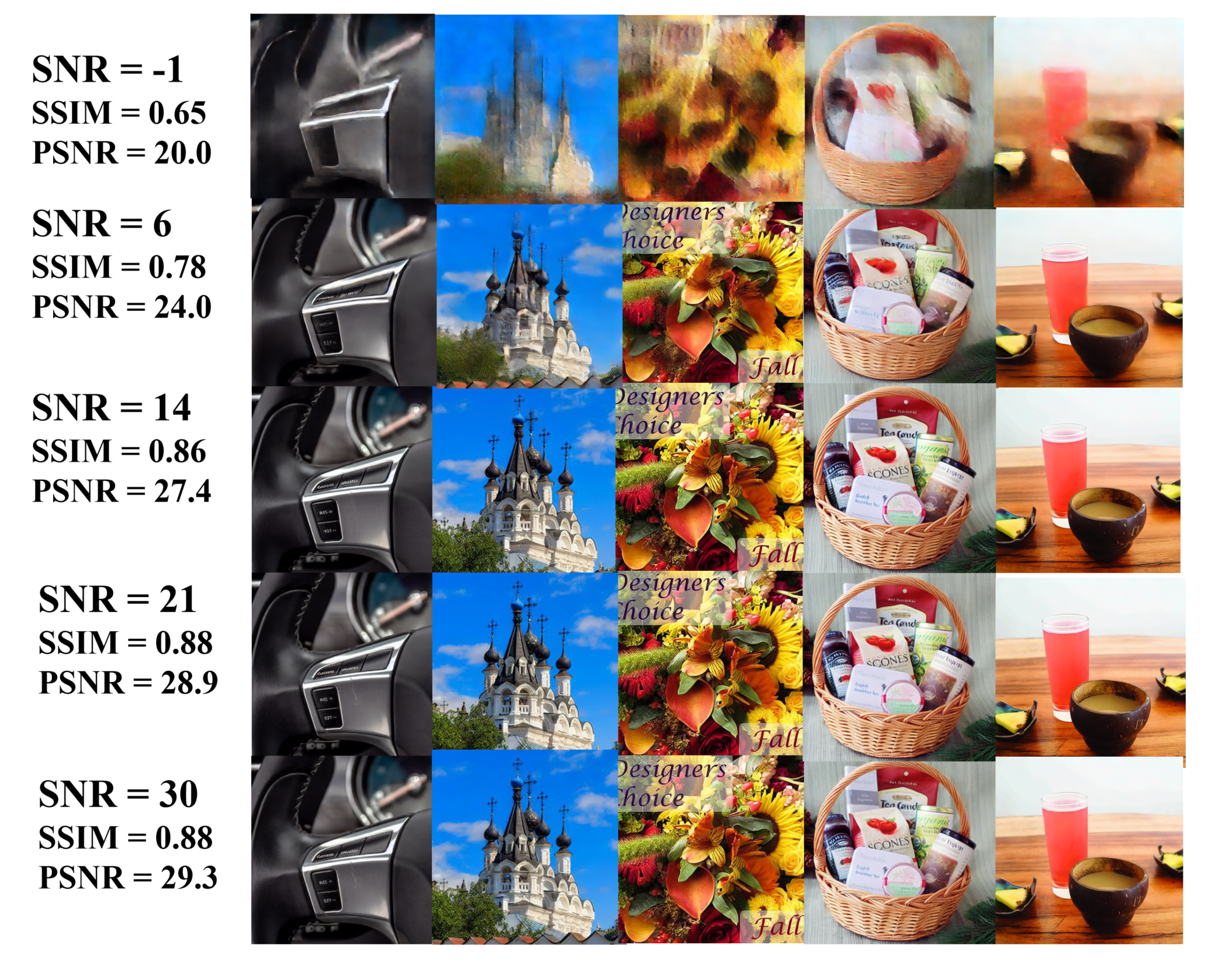}
\caption{Examples of reconstructed images produced by the proposed Latent-Diff DNSC scheme.}
\label{1}
\vspace{-1em}
\end{figure}

PSNR and SSIM\cite{5596999} are used as metrics, representing image quality and similarity, respectively. Fig. \ref{1} illustrates the visualization of a subset of images recovered under different SNRs using the proposed Latent-Diff DNSC scheme with a compression ratio $r_s$ of $1/48$.

\begin{figure}[t]
\centering
\includegraphics[width=0.47\textwidth,height=0.35\textwidth]{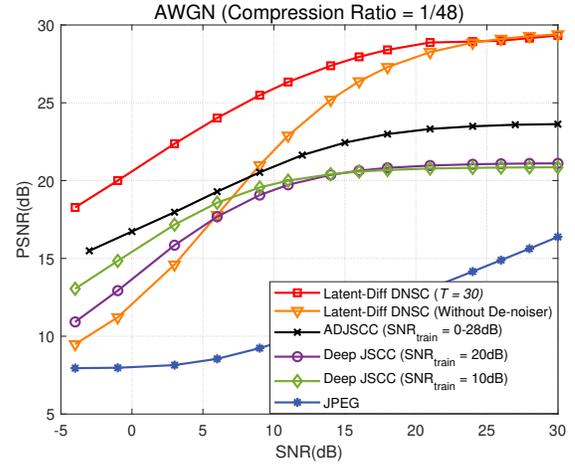}
\caption{PSNR of the proposed Latent-Diff DNSC scheme and baseline schemes on LAION2B-EN images versus different SNRs.}
\label{PSNR}
\end{figure}

Fig. \ref{PSNR} presents the PSNR performance of the proposed Latent-Diff DNSC scheme and four baselinse schemes. The PSNR performance of the proposed Latent-Diff DNSC scheme is significantly better than all baseline schemes over different SNRs. Specifically, at high SNRs, DeepJSCC performs better when trained with SNR of 20 dB compared to SNR of 10 dB, while at low SNRs, DeepJSCC performs better when trained with SNR of 10 dB compared with SNR of 20 dB. The reason is that DeepJSCC has difficulty in online capturing varying channel features under the fixed offline training SNR. ADJSCC addresses this issue and has better performance than JSCC, whether trained with SNR of 10 or 20 dB. Furthermore, our proposed Latent-Diff DNSC scheme, with its superior ability to capture channel characteristics and de-noising capability, outperforms ADJSCC. Under low SNRs, the Latent-Diff DNSC improves PSNR by approximately 3 dB compared to ADJSCC. As the SNR increases, this advantage becomes more pronounced, with the PSNR improvement of approximately 5.7 dB at the SNR of 30 dB. Additionally, the Latent-Diff DNSC scheme without the de-noiser performs poorly at low SNRs. It can be explained that the absence of added noises during the first training stage prevents the encoder and decoder from learning accurate distribution of channel properties under various SNRs. However, it performs well at high SNRs, verifying the excellent image reconstruction ability of the Latent-Diff DNSC scheme under high $r_s$.

\begin{figure}[t]
\centering
\includegraphics[width=0.47\textwidth,height=0.35\textwidth]{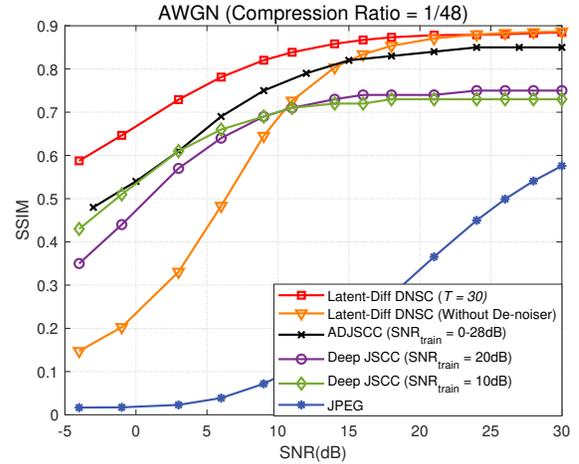}
\caption{SSIM of the proposed Latent-Diff DNSC scheme and baseline schemes on LAION2B-EN images versus different SNRs.}
\label{SSIM}
\end{figure}

Fig. \ref{SSIM} provides the SSIM performance of the proposed Latent-Diff DNCS scheme and baseline schemes versus different SNRs. Similar with Fig. \ref{PSNR}, the proposed Latent-Diff DNSC scheme outperforms all baseline schemes in different channel conditions. Furthermore, under low SNRs, the SSIM of the proposed Latent-Diff DNSC is approximately 0.1 higher than that of ADJSCC. AS the SNR decreases, this advantage becomes even more pronounced.

\subsection{De-noising Trade-off Between Semantic and Noise Components}
\begin{figure}[t]
\centering
\includegraphics[width=0.47\textwidth,height=0.35\textwidth]{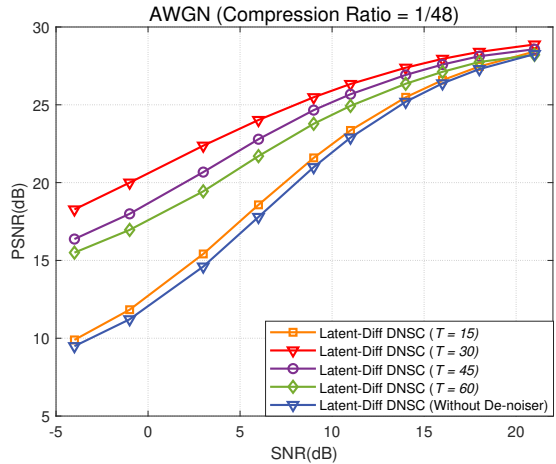}
\caption{PSNR of the proposed Latent-Diff DNSC scheme on LAION2B-EN images versus different SNRs.}
\label{PSNR_DNSC}
\end{figure}

\begin{figure}[t]
\centering
\includegraphics[width=0.47\textwidth,height=0.35\textwidth]{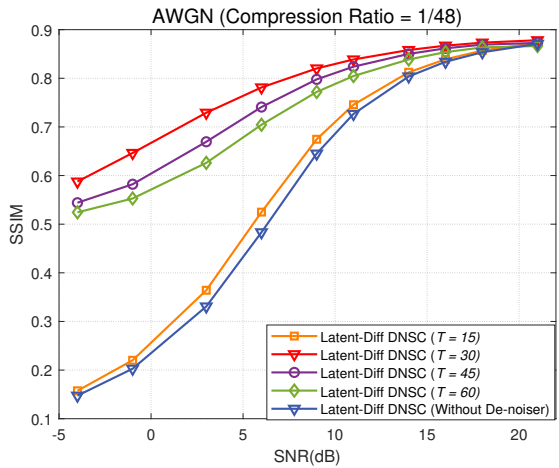}
\caption{SSIM of the proposed Latent-Diff DNSC scheme on LAION2B-EN images versus different SNRs.}
\label{SSIM_DNSC}
\end{figure}

Fig. \ref{PSNR_DNSC} and \ref{SSIM_DNSC} show the performance variations of the Latent-Diff DNSC system in terms of PSNR and SSIM indicators under different de-noising steps. When the de-noising steps are 15, the system performance is severely degraded, indicating that a large amount of noises in semantic vectors are still not eliminated. When the de-noising steps are 45 and 60, the system performance is lower than that of the system with de-noising steps of 30. This is due to the fact that in the iterative de-noising process, each de-noising step subtracts the estimated noise component from the received semantic vector. Excessive de-noising steps may lead to removal of semantic vector components that do not belong to the noise component, resulting in a decrease in system performance.

\section{Conclusions}

In this paper, we propose the DNSC system to combat channel noises by introducing the semantic de-noiser. Based on the DNSC framework, we further design the Latent-Diff DNSC scheme, where the VAE serves as the encoder and decoder, and the semantic de-noiser is achieved by a diffusion process in the latent space involved VAEs and adversarial learning. In this way, chanel noises are modeled in a diffusion manner and gradually eliminated by variational inference. Such design can enable adaptive online de-noising under different SNRs estimated by an SNR estimator. Experimental results on the open-source datasets show that the proposed Latent-Diff DNSC approach can achieve better transmission performance in PSNR and SSIM than four baseline schemes with high compression ratio of $1/48$. Therefore, the proposed method is a promising solution for image semantic communications under dynamic channel environments in future 6G systems.


\begin{thebibliography}{10}
\providecommand{\url}[1]{#1}
\csname url@samestyle\endcsname
\providecommand{\newblock}{\relax}
\providecommand{\bibinfo}[2]{#2}
\providecommand{\BIBentrySTDinterwordspacing}{\spaceskip=0pt\relax}
\providecommand{\BIBentryALTinterwordstretchfactor}{4}
\providecommand{\BIBentryALTinterwordspacing}{\spaceskip=\fontdimen2\font plus
\BIBentryALTinterwordstretchfactor\fontdimen3\font minus
  \fontdimen4\font\relax}
\providecommand{\BIBforeignlanguage}[2]{{%
\expandafter\ifx\csname l@#1\endcsname\relax
\typeout{** WARNING: IEEEtran.bst: No hyphenation pattern has been}%
\typeout{** loaded for the language `#1'. Using the pattern for}%
\typeout{** the default language instead.}%
\else
\language=\csname l@#1\endcsname
\fi
#2}}
\providecommand{\BIBdecl}{\relax}
\BIBdecl

\bibitem{saad2019vision}
W.~Saad, M.~Bennis, and M.~Chen, ``A vision of 6g wireless systems:
  Applications, trends, technologies, and open research problems,'' \emph{IEEE
  network}, vol.~34, no.~3, pp. 134--142, 2019.

\bibitem{yang2022semantic}
W.~Yang, H.~Du, Z.~Q. Liew, W.~Y.~B. Lim, Z.~Xiong, D.~Niyato, X.~Chi, X.~S.
  Shen, and C.~Miao, ``Semantic communications for future internet:
  Fundamentals, applications, and challenges,'' \emph{IEEE Communications
  Surveys \& Tutorials}, 2022.

\bibitem{yang2022semantic2}
W.~Yang, Z.~Q. Liew, W.~Y.~B. Lim, Z.~Xiong, D.~Niyato, X.~Chi, X.~Cao, and
  K.~B. Letaief, ``Semantic communication meets edge intelligence,'' \emph{IEEE
  Wireless Communications}, vol.~29, no.~5, pp. 28--35, 2022.

\bibitem{dong2022semantic}
C.~Dong, H.~Liang, X.~Xu, S.~Han, B.~Wang, and P.~Zhang, ``Semantic
  communication system based on semantic slice models propagation,'' \emph{IEEE
  Journal on Selected Areas in Communications}, vol.~41, no.~1, pp. 202--213,
  2022.

\bibitem{o2016learning}
T.~J. O'Shea, K.~Karra, and T.~C. Clancy, ``Learning to communicate: Channel
  auto-encoders, domain specific regularizers, and attention,'' in \emph{2016
  IEEE International Symposium on Signal Processing and Information Technology
  (ISSPIT)}.\hskip 1em plus 0.5em minus 0.4em\relax IEEE, 2016, pp. 223--228.

\bibitem{xie2021deep}
H.~Xie, Z.~Qin, G.~Y. Li, and B.-H. Juang, ``Deep learning enabled semantic
  communication systems,'' \emph{IEEE Transactions on Signal Processing},
  vol.~69, pp. 2663--2675, 2021.

\bibitem{weng2021semantic}
Z.~Weng and Z.~Qin, ``Semantic communication systems for speech transmission,''
  \emph{IEEE Journal on Selected Areas in Communications}, vol.~39, no.~8, pp.
  2434--2444, 2021.

\bibitem{bourtsoulatze2019deep}
E.~Bourtsoulatze, D.~B. Kurka, and D.~G{\"u}nd{\"u}z, ``Deep joint
  source-channel coding for wireless image transmission,'' \emph{IEEE
  Transactions on Cognitive Communications and Networking}, vol.~5, no.~3, pp.
  567--579, 2019.

\bibitem{du2023ai}
H.~Du, J.~Wang, D.~Niyato, J.~Kang, Z.~Xiong, and D.~I. Kim, ``Ai-generated
  incentive mechanism and full-duplex semantic communications for information
  sharing,'' \emph{arXiv preprint arXiv:2303.01896}, 2023.

\bibitem{hu2022robust}
Q.~Hu, G.~Zhang, Z.~Qin, Y.~Cai, G.~Yu, and G.~Y. Li, ``Robust semantic
  communications with masked vq-vae enabled codebook,'' \emph{arXiv preprint
  arXiv:2206.04011}, 2022.

\bibitem{xu2021wireless}
J.~Xu, B.~Ai, W.~Chen, A.~Yang, P.~Sun, and M.~Rodrigues, ``Wireless image
  transmission using deep source channel coding with attention modules,''
  \emph{IEEE Transactions on Circuits and Systems for Video Technology},
  vol.~32, no.~4, pp. 2315--2328, 2021.

\bibitem{schuhmann2022laion}
C.~Schuhmann, R.~Beaumont, R.~Vencu, C.~Gordon, R.~Wightman, M.~Cherti,
  T.~Coombes, A.~Katta, C.~Mullis, M.~Wortsman \emph{et~al.}, ``Laion-5b: An
  open large-scale dataset for training next generation image-text models,''
  \emph{arXiv preprint arXiv:2210.08402}, 2022.

\bibitem{rombach2022high}
R.~Rombach, A.~Blattmann, D.~Lorenz, P.~Esser, and B.~Ommer, ``High-resolution
  image synthesis with latent diffusion models. 2022 ieee,'' in \emph{CVF
  Conference on Computer Vision and Pattern Recognition (CVPR)}, 2022, pp.
  10\,674--10\,685.

\bibitem{5596999}
A.~Horé and D.~Ziou, ``Image quality metrics: Psnr vs. ssim,'' in \emph{2010
  20th International Conference on Pattern Recognition}, 2010, pp. 2366--2369.

\end{thebibliography}

\bibliographystyle{IEEEtran}
\end{document}